\documentstyle[aps,epsfig]{revtex}

\begin{document}

\preprint{MA/UC3M/22/96}



\title{Noise Effects on Synchronized Globally Coupled Oscillators} 

\author{Esteban Moro$^*$ and Angel S\'anchez$^{\dag}$}

\address{Departamento de Matem\'aticas and Grupo Interdisciplinar de Sistemas
Complicados,\\ 
Universidad Carlos III de Madrid, E-28911 Legan\'{e}s, Madrid, Spain}

\date{\today}

\maketitle

\begin{abstract}
The synchronized phase of globally coupled nonlinear oscillators 
subject to noise fluctuations is studied by means of a new 
analytical approach able to tackle general couplings, nonlinearities,
and noise temporal correlations. Our results show that the interplay
between coupling and noise modifies the effective frequency 
of the system in a non trivial way. Whereas for linear couplings
the effect of noise is always to increase the effective 
frequency, for nonlinear couplings the noise influence is shown 
to be positive or negative depending on the problem parameters.
Possible experimental verification of the results is discussed.
\end{abstract}


Systems of coupled nonlinear oscillators are a generic paradigm 
of a whole class of problems arising in physics, chemistry, 
or biology \cite{Kuramoto}. Examples are Josephson Junction 
arrays (JJA) \cite{JJA}, charge-density waves \cite{Fukuyama}, thin 
film fabrication \cite{CW}, chemical reactions and cardiac tissue
\cite{Winfree}, neuronal activity \cite{SGK}, and many more (see, 
e.g., Ref.\ \cite{Kuramoto} and references therein). Due to 
the diversity of individual oscillators or to external (thermal) 
noise, these systems generally exhibit 
cooperative dynamical response or incoherent behavior as a function 
of the relevant parameters. The study of the transition between
both regimes and the nature of the so-called {\em synchronized} phase 
requires the calculation of the phase
distribution probability density. This involves solving
a highly nonlinear, partial differential (Liouville) equation which is
almost always a formidable task and only perturbative results can be
obtained for practically all models of interest. 

In this paper, we approach these problems by introducing a 
new method, from a completely different viewpoint, that avoids 
dealing with partial differential equations at all. We focus on
the stochastic (Langevin) evolution equation that allows
to perturbatively 
obtain accurate results in a very simple and direct manner. This 
method can be applied to any system of $N$ coupled homogeneous 
oscillators evolving 
in the presence of random forces, i.e., systems of the form 
\begin{equation}
\label{model}
\dot \phi_i = \omega - f(\phi_i) + \frac{\kappa}{N} 
\sum_{j=1}^{N} \Gamma(\phi_i - \phi_j) + \sigma \eta_i(t),
\end{equation}
where $\phi_i$ is the phase of oscillator $i$, $\omega > 0$ is the 
oscillator frequency, $\kappa$ is a constant, $f(\phi)$ is any 
2$\pi$-periodic function, 
$\Gamma$ is any separable (see below)
function, and the random term $\eta_i$ is local (uncorrelated from 
site to site), Gaussian and stationary. Aside from these 
requirements, $\eta_i$ can be any process defined by a stochastic differential 
equation, e.g., it can be white or colored noise. 
To illustrate our method, we 
consider an Ornstein-Uhlenbeck process, given by $\dot \eta_i = 
-\gamma \eta_i + \gamma \xi_i(t)$, with $\xi_i(t)$ being uncorrelated
Gaussian
white noises [$\langle \xi_i(t) \xi_j(t')\rangle = \delta_{ij} 
\delta(t-t')$; 
$\langle \cdots \rangle$ stands for averages over noise realizations]
and $\tau = 1/\gamma$ being the correlation time.

Far from being academic, model (\ref{model}) contains already several 
important applications.
To begin with, the coupling can describe 
many different systems: For instance, in the case of 
the phase approximation of globally coupled 
nonlinear oscillators or of a JJA, 
$\Gamma(\phi_i - \phi_j) = \sin(\phi_i-\phi_j + \alpha)$, $\alpha$ 
being a constant, whereas in 
reaction-diffusion (RD) problems such as those arising in growth models
\cite{CW},
the coupling comes from the 
discretization of some spatial linear operator followed by a mean field
approximation, yielding $\Gamma(\phi_i - \phi_j) = \phi_i -\phi_j$. Second,
the nonlinear term, $f(\phi)$, for the oscillator can be chosen as
required in each problem; as in most
coupled oscillator models, we take here $f(\phi)=\sin\phi$, but our 
method can be applied to other choices as well. 

We now begin the study of Eq.\ (\ref{model}) in the case when,
as mentioned above, $\Gamma$ is separable, i.\ e., $\Gamma(\phi_j -
 \phi_i) = \sum_n A^n(\phi_i) B^n(\phi_j)$. This 
technicality is needed to simplify the study but does not
not restrict much the applicability of our calculations. Eq.\ (\ref{model})
becomes then
\begin{equation}
\label{modelb}
\dot \phi_i = \omega - f(\phi_i) + \frac{\kappa}{N}
\sum_{n} A^n(\phi_i)\left[\sum_{j=1}^N B^n(\phi_j)\right] + \sigma \eta_i(t).
\end{equation}
The first step is to take the limit $N\to\infty$. In this limit
the sum within the brackets above can be 
computed in terms of the mean value of
$\phi$ \cite{mean}. As way of example, we focus on the 
RD problems and JJA's, for which we arrive (respectively) at
\begin{eqnarray}
\label{lineal}
\dot \phi&=&\omega - a \sin \phi + 
\kappa \left[ \langle \phi \rangle - \phi\right] + \sigma \eta(t), \\
\label{kura}
\dot \phi&=&\omega - a \sin \phi + \nonumber\\
& & + \kappa \left[  
\cos (\phi -\alpha) \langle\sin\phi\rangle -
\sin (\phi -\alpha) \langle\cos\phi\rangle \right] + \sigma \eta(t).
\end{eqnarray}
We note
that we have thus reduced the $N$ equations in (\ref{model}) to a 
single (self-consistent) one. 

We deal first with eq.\ (\ref{lineal}) for RD problems, hereafter called {\em linear 
coupling model}. When $\sigma=0$, it can be straightforwardly shown that
if $\omega \leq \omega_C = a$ there is a 
stable time independent pinned solution $\phi = 
\sin^{-1} \omega/a$, whereas if $\omega > \omega_C$ the phase increases
oscillatorily in time with frequency
$\Omega^2\equiv\omega^2 - a^2$. The natural question to ask is whether 
this scenario, involving a {\em depinning 
transition} at $\omega_C$, changes when noise is switched on, and if 
so, how. To address this issue we 
choose $\langle \phi \rangle$ as our order parameter, because 
in the pinned phase $\langle \phi 
\rangle$ is constant whereas in the depinned phase it increases in 
time. For noise intensities which are 
small compared to $\omega$ or $\kappa$, we expand $\omega_C$
and $\phi$ in powers of $\sigma$, $\omega_C = a + a^{(1)} 
\sigma + a^{(2)} \sigma^2 + {\cal O}(\sigma^3)$ and $\phi = \phi^{(0)} + \sigma 
\phi^{(1)} + \sigma^2 \phi^{(2)} + {\cal O}(\sigma^3)$. 
We first look for possible changes in $\omega_C$, by inserting these two 
expressions in Eq.\ (\ref{lineal}); collecting powers of $\sigma$ and 
imposing that $\langle\dot{\phi}^{(i)}\rangle=0$ for $i=0,1,\ldots$ (i.e.,
that the oscillators remained pinned), we 
can compute the $a^{(i)}$ corrections to $\omega_C$. The final result is
\begin{eqnarray}
\frac{\omega_C}{a}&=&1 - \frac{\sigma^2}{4 \kappa (\kappa \tau + 1)} + \nonumber\\
& & +\frac{\sigma^4}{32 \kappa^4} \left[ \frac{\kappa^2 - a^2}{(\kappa \tau +1)^2}
+{\cal O}(\tau) \right] + {\cal O}(\sigma^6).
\end{eqnarray}

From this expression, we immediately see first, that the transition occurs
at a value of $\omega$ which is lower 
than in the deterministic case, which means 
that the noise is actually helping the oscillators to overcome the barrier
and start their motion. Another important conclusion is that for a given, 
fixed set of the other parameters, $\tau$ controls the state of the 
system: whereas for low $\tau$ values, the oscillators are depinned, 
as $\tau$ increases the pinned phase is set on. We thus see 
the importance of the correlation time in the critical region of the 
system.

Having found the changes in $\omega_C$, we
now turn to the dynamics in the depinned phase, when $\omega > 
\omega_C$. We again expand $\phi$ in powers of $\sigma$ and write down 
equations for each contribution, which read 
\begin{mathletters}
\begin{eqnarray}
\label{x0}
\langle \dot \phi^{(0)}\rangle&=&\omega - a \sin\langle\phi^{(0)}\rangle;\\
\label{x12}
\langle \dot \phi^{(1)}\rangle&=&0;\\
\langle (\dot\phi^{(1)})^2\rangle&=
&-2 \langle (\phi^{(1)})^2\rangle (\kappa + a \cos\langle\phi^{(0)}\rangle) + 2 \zeta(t), \\
\dot \zeta&=&
\frac{\gamma}{2} - \zeta (\kappa + \gamma + a \cos\langle\phi^{(0)}\rangle);\\
\label{x2}
\langle \dot \phi^{(2)} \rangle&=& -
a \langle \phi^{(2)} \rangle \cos\langle\phi^{(0)}\rangle
+ a \frac{\langle (\phi^{(1)})^2\rangle}{2} \sin\langle\phi^{(0)}\rangle.
\end{eqnarray}
\end{mathletters}
where $\zeta \equiv \langle \phi^{(1)} \eta \rangle$.
Eq.\ (\ref{x2}) implies that $\langle \phi^{(2)} \rangle$ 
grows in time faster than $\langle \phi^{(0)} \rangle$, and therefore our 
expansion is not correct. This problem, very well 
known in deterministic equations \cite{kevor}, can be cured by realizing 
that the system dynamics does not depend only on one time scale but on two, 
$t$ and $t^* = \mu(\sigma)t$. Among the different approaches one can use to
deal with this, we choose Linstedt's method: we assume that 
$\phi^{(i)}$ depend on time through the combination $t + t^*$; as for 
$\mu(\sigma)$, dimensional analysis shows that the noise term is of 
order one when $\mu(\sigma) = \nu \sigma^2$. 
Introducing this new time scale in equations (\ref{x0}-\ref{x2}) and
imposing the usual solvability condition on equation (\ref{x2}) (Fredholm's
alternative, see e.\ g.\ \cite{kevor} for details) we obtain
\begin{equation}
\label{nu}
\nu = \frac{a}{2T} \int_0^T
\frac{\langle \phi_1^2 \rangle \sin\langle \phi_0 \rangle}
{\omega - a \sin\langle \phi_0 \rangle} dt ,
\end{equation}
where $T = 2\pi / \Omega$
is the period of $\langle\dot{\phi}^{(0)}
\rangle$ computed from eq.\ (\ref{x0}).
With this value, the effective frequency of the oscillations, defined
as $\omega_{\rm eff} = \lim_{t,s\to\infty} \frac{1}{s}[\langle{\phi}(t+s)\rangle - 
\langle{\phi}(t)\rangle ]$, is:
\begin{equation}
\label{weff}
\omega_{\rm eff} = \Omega (1 + \nu \sigma^2 + {\cal O}(\sigma^4))
\end{equation}
\begin{figure}
\begin{center}
\vspace{-.2cm}
\epsfig{file=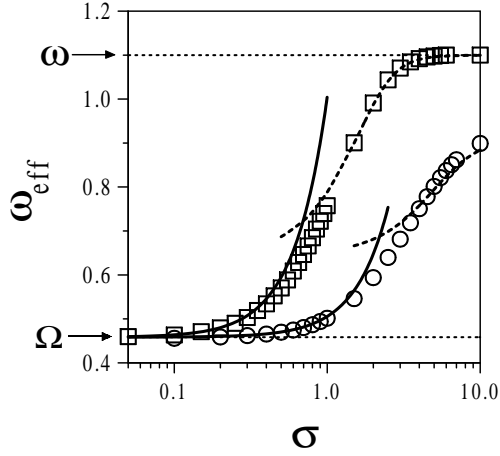,width=2.8in, clip=}
\caption[]{{\small Predictions given by eqns.\ 
(\ref{weff}) up to order $\sigma^2$ (solid line) and (\ref{weff1})
up to order $a^2$ (dashed line) are shown
with the numerical solution of eqns.\ 
(\ref{lineal}) for $N=5\,000$, $a=1$, $\kappa =1$, $\omega = 1.1$ and 
$\tau=0$ (squares) or $\tau=10$ (circles).
The numerical calculation was done with an explicit 2.0 order weak 
scheme \cite{num}.}}
\label{fig1}
\end{center}
\end{figure}
Figure \ref{fig1} compares our analytical, 
order $\sigma^2$ expression (\ref{weff}) for $\omega_{\rm eff}$ 
with the numerical solution of Eqs.\ 
(\ref{lineal}) up to $\sigma\sim 0.5$, already a not so small value.
The expression for $\omega_{\rm eff}$ above indicates that 
the noise increases the frequency as compared with the deterministic
value, in agreement with our previous conclusion that the noise helps
the oscillators jump over the potential barrier: The larger the noise
strength, the easier the potential barrier to overcome, effectively
suppresing (renormalizing) it in the $\sigma\to\infty$ limit.
The fact that $ \nu > 0$, implying that the frequency increases
with the noise strength, can be proven rigorously when $\tau=0$, i.\ e., for
white noise, but we conjecture that this result is general for linearly
coupled systems. 

Another advantage of our approach is that it can be applied to other limits,
such as the case when the 
noise intensity is larger than $\omega$ and $a$. 
Expanding the solution now in powers of $a$, the same steps as before yield,
up to order $a^2$,  
\begin{eqnarray}
\label{weff1}
& & \omega_{\rm eff}=\omega - \displaystyle{a^2\over2} 
\exp\left(-\frac{\tilde\sigma^2}{2\kappa}\right)\Bigg[ 
\frac{\kappa^2 \omega}{\kappa^2 \omega^2 + \omega^4} +
\nonumber\\
& & 
\displaystyle{\int}_0^{\infty} \exp \left\{ 
{\frac{\tilde\sigma^2(\kappa\tau e^{-s/\tau} - 
e^{-\kappa s})}{2\kappa(\kappa\tau-1)} - \kappa s}
\right\} \sin \omega s \ d s \Bigg]
\end{eqnarray}
where $\tilde\sigma = \sigma/\sqrt{\kappa \tau +1}$. Again, the comparison 
with the numerical solution (see Fig.\ \ref{fig1}) 
shows that our approach is very accurate, in 
this case for values of $\sigma$ down to $\sigma\sim 1.2$, very close to the 
values of $\omega$ and $a$. We want to stress that, except for a small interval 
$0.5\leq \sigma\leq 1.2$, our analytical predictions (\ref{weff}) and 
(\ref{weff1}) describe to a high degree 
of accuracy the numerical solution for any noise intensity, 
providing a global picture of the main features of the linearly-coupled 
system behavior. 
\begin{figure}
\begin{center}
\epsfig{file=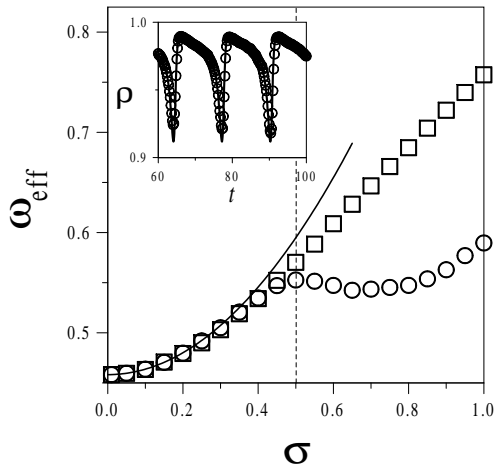, width=2.8in, clip=}
\caption[]{Comparison of the numerical
values of $\omega_{\rm eff}$ for the linear (squares) and KS
(circles) models with the small $\sigma$ 
approximation (\ref{weff}) (solid lines). Parameters are as in 
Fig.\ \ref{fig1} with $\tau=0$. Note the 
synchronization-desynchronization transition for the KS model is at 
$\sigma \approx 0.5$.
The inset shows the computed time evolution of the order parameter,
$\rho$, (see text)
in the KS model for $\sigma = 0.2$ (circles) and the comparison with our 
analytical aproximation (solid line).}
\label{fig2}
\end{center}
\end{figure}

We now move on to the nonlinear problem (\ref{kura}) in
the case $\alpha=0$; Eqs.\ (\ref{model}) are then known as
{\em Kuramoto-Sakaguchi model} (KS). In the small 
noise regime, our technique applied to the KS model 
leads once again to Eqs.\ (\ref{x0}-\ref{x2}). This stems from the
fact that, when $\sigma \rightarrow 0$, the phases of individual oscillators are 
similar, and consequently $\sin(\phi_j-\phi_i) \approx 
\phi_j-\phi_i$. However, due to the periodicity of the KS coupling term,
the model is invariant under changes 
$\phi_i-\phi_j \leftrightarrow \phi_i - \phi_j + 2 n \pi$, for any integer $n$.
Therefore, $2\pi$
jumps between oscillators are possible without consequence other
than the breakdown of the linear approximation. Of course, 
the larger the noise strength, the more likely such jumps are, and
the KS model becomes desynchronized at a finite value of $\sigma$, i.\ e., 
it has a true synchronization-desynchronization transition 
at $\sigma^2 = \kappa$ when $a=0$ \cite{Kuramoto}, and at lower values 
for $a \not= 0$ \cite{nota2}.
 Up to that point, it can be seen from Fig.\ \ref{fig2} 
that the linear model, the KS model and our analytical prediction 
Eq.\ (\ref{weff}) are all in excellent agreement with each other. For our
parameters, $a = 1$, $\omega = 1.1$, the transition takes place at 
$\sigma \approx 0.5$. 
\begin{figure}
\begin{center}
\epsfig{file=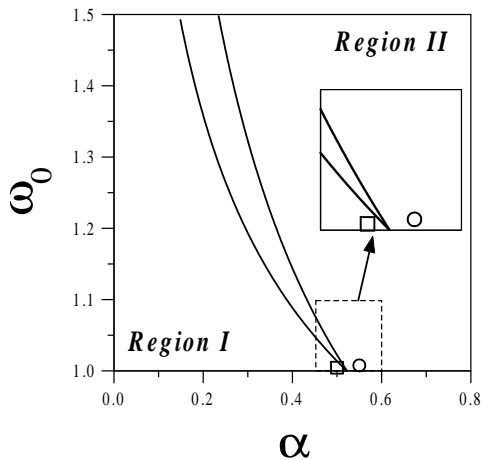, width=2.8in, clip=}
\caption[]{Diagram of the possible signs for $\nu(\alpha)$ for different
values. Leftmost line
corresponds to $\tau=10$, rightmost one to $\tau=0$. In both cases, 
Region I ($\nu_{\alpha} > 0$) is ond the left of the line, Region II 
($\nu_{\alpha} < 0$) lies on the right. The square and the circle
points correspond to the cases analyzed in figure (\ref{fig4}).}
\label{fig3}
\end{center}
\end{figure}
The order parameter for the synchronization-desynchronization transition is 
$\rho = \langle \cos \phi \rangle^2 + \langle \sin \phi \rangle^2$; in the
synchronized phase, $\rho$ oscillates around a nonzero value, while in the
desynchronized phase, $\rho = 0$. 
The inset in Fig.\ \ref{fig2} compares the numerical value of $r$ with the 
result of our aproximation, $\rho = 1 - \sigma^2 \langle (\phi^{(1)})^2 \rangle +
{\cal O}(\sigma^3)$, exhibiting the quantitative validity of our approach.
The agreement is very good up to the critical value of $\sigma$. As for 
the limit $\sigma\to\infty$, the computations, although feasible, are
quite involved, and therefore we do not present the results here.

\begin{figure}
\begin{center}
\epsfig{file=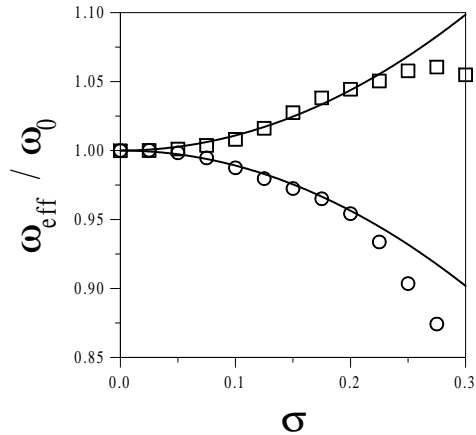, width=2.8in, clip=}
\caption[]{{\small $\omega_{\rm eff}$ obtained from simulations 
for $\omega_0= 1.006$, 
$\tau = 0$, $\kappa = 1$ and $\alpha=0.5$ (squares) or
$\alpha = 0.55$ (circles), 
compared to the analytical approximations (solid lines)
given by eqns.\ (\ref{weff}) and (\ref{nualfa}). [Squares and
circles refer to the corresponding points in Fig.\ \ref{fig3}].}}
\label{fig4}
\end{center}
\end{figure}

In the two cases analyzed so far, in the synchronized phase 
noise helps the oscillators overcome 
the nonlinear potential, hence effectively increasing their frequency.  
However, this intuitively reasonable picture is 
not the generic situation: Antisymmetric couplings verifying $\Gamma(-\phi) 
= -\Gamma(\phi)$ keep the difference between oscillators small, forcing 
their motion to be approximately the same, and then the above picture holds,
but if the coupling lacks this symmetry (like 
$\Gamma(\phi_i-\phi_j+\alpha)$ for nonzero $\alpha$), 
the oscillators do not tend to 
synchronize (they rather have phases separated by a factor $\alpha$).
In the latter case, 
the noise competes with the 
frustration induced by the coupling and its activation effect disappears.
Our calculations for Eqs.\ (\ref{kura}) when $\alpha\neq 0$ and 
$\sigma \to 0$ show that this is indeed what occurs. Our method leads now
to a 
value of $\nu$, the $\sigma^2$ correction to the frequency,
which depends on $\alpha$, given by 
\begin{equation}
\label{nualfa}
\nu_{\alpha} = \frac{1}{2T} \int_0^T \frac{\langle \phi_1^2 \rangle
(a \sin \langle \phi_0 \rangle
-2 \kappa \sin\alpha)}
{\omega_0 -a\sin\langle \phi_0 \rangle}\  dt
\end{equation}
where $\omega_0 \equiv \omega + \kappa \sin \alpha$, this parameter playing 
the role of an effective driving.
Eq.\ (\ref{nualfa}) shows that, when  
$\alpha < \alpha_0 \equiv 
\sin^{-1}(1/2\kappa)$, $\nu_{\alpha}$ is positive or negative
depending on the value of $\omega$, whereas when 
$\alpha > \alpha_0$ we have $\nu_{\alpha} < 0$. This divides 
the $(\omega,\alpha)$-space in two regions, according to the possible signs 
of $\nu_\alpha$, as depicted in Fig.\ \ref{fig3}.  Interestingly, in region I
$\alpha$ (or equivalently $\omega$) acts as a switching 
parameter between noise induced acceleration or deceleration of the oscillator
motion, as illustrated in Fig.\ \ref{fig4}. Note that this is a
novel, noise intrinsic phenomenon, absent in the deterministic problem, 
whose origin is the nonlinear coupling: If we add a constant
to the coupling in Eq.\ (\ref{lineal}) only the external 
frequency is changed ($\omega \rightarrow \omega + \kappa \alpha$).

In summary, we have introduced a procedure to study systems of globally
coupled oscillators in the synchronized phase,
general enough to analyze any stationary 
random processes generated by a stochastic differential equation, any form 
of the nonlinearity, and a large family of coupling terms. We have used it
to analyze problems
relevant in a number of physical contexts, obtaining 
fairly good results in a much more straightforward way than the 
approaches proposed so far. 
For linear coupling, we have computed the noise-induced corrections to the 
critical value for the depinning transition as well as the 
effective frequency in the depinned phase for almost all values of the 
noise. We have shown that the same results hold for the KS model.
In the general case of frustrated
JJA's we have found that the effect of noise on the effective frequency is not
trivial, leading to an increase or a decrease of the 
effective frequency depending on the system parameters and the driving 
frequency $\omega$. This very important result shows the power of our 
technique as it was not known prior to our work. Our predictions can be 
directly checked by experiments in very many fields, among which we suggest
JJA's subject to a magnetic field where tuning the voltage for a given 
temperature one should be able to find the two different behaviors predicted.

We are indebted to J. A. Cuesta and R. Cuerno for discussions.
Work at GISC (Legan\'es) was supported by CICyT (Spain) grant MAT95-0325 and
DGES (Spain) grant PB96-0119.


\vspace*{-0.5cm}
\begin{references}
\vspace*{-1.5cm}
\bibitem[*]{eme} Electronic address: {\tt eme@math.uc3m.es}
\bibitem[\dag]{anxo} Electronic address: {\tt anxo@math.uc3m.es}

\bibitem{Kuramoto} Y.\ Kuramoto, {\em Chemical Oscillations, Waves, and 
Turbulence} (Springer, Berlin, 1984); S.\ H.\ Strogatz, in {\em Frontiers
in Mathematical Biology}, edited by S.\ Levin (Springer, Berlin, 1994). 

\bibitem{JJA} A. V. Ustinov {\em et al.}, Phys.
Rev. B {\bf 47}, 8357 (1993);
S. Watanabe {\em et al.}, Physica D {\bf 97}, 429 (1996).

\bibitem{Fukuyama} H.\ Fukuyama, J.\ Phys.\ Soc.\ Jpn.\ {\bf 41}, 513 
(1976); {\bf 45}, 1474 (1978); 
H.\ Fukuyama and P.A. Lee, Phys.\ Rev.\ B {\bf 17}, 535 (1977);
P.\ A.\ Lee and T.\ M.\ Rice, Phys.\ Rev.\ B {\bf 19}, 3970 (1979).

\bibitem{CW} J.\ D.\ Weeks and G.\ H.\ Gilmer, Adv.\ Chem.\ Phys.\ {\bf 40},
157 (1979);
A.\ S\'anchez {\em et al.}, Phys.\ Rev.\ B {\bf 51}, 14\,664 (1995);
E.\ Moro {\em et al.}, Phys.\ Rev.\ Lett.\
{\bf 78}, 4982 (1997).

\bibitem{Winfree} A.\ T.\ Winfree, {\em When Time Breaks Down}
(Princeton University, Princeton, 1987). 

\bibitem{sto} C. W. Gardiner, {\em Handbook of Stochastic Methods}
(Springer-Verlag, Berlin, 1996).

\bibitem{SGK} H.\ Sompolinsky {\em et al.}, Phys.\ Rev.\ A {\bf 43}, 6990 (1991).

\bibitem{mean} When $N\to\infty$, the average of the phase $\phi_i$ over the
ensemble of oscillators equals its average over realizations of the noise 
$\eta_i(t)$. 


\bibitem{kevor} J. Kevorkian and J. D. Cole, {\em Multiple Scale and Singular
Perturbations Methods} (Springer, New-York, 1996). 

\bibitem{num} P. E. Kloeden and E. Platen, {\em Numerical Solution 
of Stochastic Differential Equations} (Springer, Berlin, 1992).

\bibitem{nota2} S. Shinomoto and Y. Kuramoto, Prog. Theor. Phys. {\bf 75} (1986), 1105.  

\end{references}
\end{document}